\documentclass{moriond}

\usepackage[T1]{fontenc}  
\usepackage{amssymb}
\usepackage{amsfonts}
\usepackage{amsmath}
\usepackage{bm}
\usepackage{dsfont}
\usepackage{xspace}

\def\be{\begin{equation}}
\def\ee{\end{equation}}
\def\bea{\begin{eqnarray}}
\def\eea{\end{eqnarray}}

\newcommand{\smo}{\textsc{SModelS}\xspace}
\newcommand{\protomodel}{proto-model\xspace}
\newcommand{\Protomodel}{Proto-model\xspace}
\newcommand{\protomodels}{proto-models\xspace}
\newcommand{\Protomodels}{Proto-models\xspace}
\newcommand{\M}{\mathbf{M}}

\newcommand{\K}{\ensuremath{K}\xspace}
\newcommand{\SM}{\ensuremath{\mathbf{SM}}\xspace}
\newcommand{\BSM}{\ensuremath{\mathbf{BSM}}\xspace}

\def\met{$E_T^{\rm miss}$}

\newcommand{\Photo}{}

\begin{document}
\vspace*{4cm}
\title{Artificial proto-modelling with simplified-model results from the LHC}

\author{Sabine Kraml$^1\,$\footnote{Speaker}, Andr\'e Lessa$^2$, Wolfgang Waltenberger$^{3,4}$}

\address{$^1$\,Laboratoire de Physique Subatomique et de Cosmologie,
Universit\'e Grenoble-Alpes,\\ CNRS/IN2P3, 53 Avenue des Martyrs, F-38026
Grenoble, France\\
$^2$\,Centro de Ci\^{e}ncias Naturais e Humanas, Universidade
Federal do ABC, Santo Andr\'e, SP\,-\,Brasil\\
$^3$\,Institut f\"ur Hochenergiephysik,  \"Osterreichische Akademie
der  Wissenschaften,\\ Nikolsdorfer Gasse 18, A-1050 Wien, Austria\\
$^4$\,University of Vienna, Faculty of Physics, Boltzmanngasse 5,
A-1090 Wien, Austria}

\maketitle\abstracts{We present a novel approach to identify potential dispersed
signals of new physics in the slew of published LHC results. It employs
a random walk algorithm to introduce sets of new particles, dubbed
``proto-models'', which are tested against simplified-model results from
ATLAS and CMS searches for new physics by exploiting the SModelS software
framework. A combinatorial algorithm identifies the set of analyses
and/or signal regions that maximally violates the Standard Model hypothesis, while
remaining compatible with the entirety of LHC constraints in our
database. 
Crucial to the method is the ability to construct a reliable likelihood in \protomodel
space; we explain the various approximations which are needed depending
on the information available from the experiments, and how
they impact the whole procedure.}

\section{Introduction}

Searches for new physics at the LHC are usually pursued on a channel-by-channel basis. A plethora of experimental analyses are thus performed in many different final states, and the outcomes interpreted as limits on new particles in the context of specific Beyond the Standard Model (BSM) scenarios, again for each search channel separately. The disadvantage of this kind of hypothesis testing---which is also done on the theory side by comparing signal predictions of specific BSM incarnations to the experimental limits---is that only a small part of the overall available data is used. Small effects of dispersed signals, which will show up simultaneously in different final states and/or signal regions, might easily be missed, or disregarded as statistical fluctuations.

A more global exploration of the LHC data is in order. As an attempt in this direction, we presented in \cite{Waltenberger:2020ygp} the prototype of a statistical learning algorithm that identifies potential dispersed signals in the slew of published LHC analyses, building candidate “\protomodels” from them, while remaining compatible with the entirety of LHC results. Such \protomodels may then be scrutinised further in dedicated analyses and, in case of a discovery, help to eventually unravel the concrete underlying BSM theory. The ultimate goal is a data-driven bottom-up approach to the quest of new physics with minimal theoretical bias. Being easily extendable, it should eventually also allow one to fold in additional information from other (future) experiments to continuously improve the picture of what data is telling us about BSM physics.


\section{The walker algorithm}

At present, our statistical learning algorithm is based on the concept of simplified models, exploiting the \smo~\cite{Kraml:2013mwa,Ambrogi:2017neo,Ambrogi:2018ujg,Alguero:2020buz} software framework and its large database of experimental results. It employs a Markov Chain Monte Carlo (MCMC)-type random walk through \protomodel space, adding or removing new particles, and randomly changing their cross sections and branching ratios. This is coupled to a combinatorial algorithm, which identifies the set of analyses and signal regions that maximally violates the Standard Model (SM) hypothesis. 

The algorithm to comb through \protomodel parameter space in order to identify the models that best fit the data is dubbed the {\bf\em walker}. It is composed of several building blocks, or “machines” that interact with each other in a well-defined fashion:
\begin{enumerate}
\item Starting with the Standard Model, the {\bf\em builder} creates \protomodels, randomly adding or removing new particles and changing any of the \protomodel parameters.
\item The \protomodel is then passed on to the {\bf\em critic}, which checks the model against the database of simplified-model results to determine an upper bound on an overall signal strength ($\mu_{\rm max}$).
\item The {\bf\em combiner} identifies all possible combinations of results and constructs a combined likelihood for each subset.
\end{enumerate}

An essential aspect in our procedure is that that neither the number nor the kind of the new particles is fixed. 
Instead, the BSM particle content, the particle masses, production cross sections and decay branching ratios (BRs) are taken as free parameters of the \protomodels.  
This implies a parameter space of varying dimensionality for the MCMC-type random walks!

\section{\Protomodel construction rules}

\Protomodels are not intended to be fully consistent theoretical models; therefore their properties are not bound by higher-level theoretical assumptions on the underlying BSM theory. Nonetheless, for practical purposes, we have to make a number of assumptions. 

Concretely, in this work we assume that all particles either decay promptly or are fully stable at detector scales. 
Moreover, since we make extensive use of simplified-model results from searches for supersymmetry (SUSY),  
we impose that  all BSM particles are odd under a $\mathcal{Z}_2$-type symmetry, so they are always pair produced and always cascade decay to the lightest state. The Lightest BSM Particle (LBP) is taken to be stable and electrically and color neutral, and hence is a dark matter candidate. Finally, only particles with masses within LHC reach are considered part of a specific \protomodel.

In the current version of the algorithm, we allow \protomodels to consist of up to 20 BSM particles:
light quark partners $X_{q}$ ($q=u,d,c,s$); heavy quark partners $X_{b}^{i}$,~$X_{t}^{i}$ ($i=1,2$); 
charged lepton and neutrino partners $X_\ell$, $X_{\nu_\ell}$ ($\ell=e,\mu,\tau$); 
a color-octet gluon partner $X_{g}$; and electroweak partners $X_{W}^{i}$,~$X_{Z}^j$ ($i=1,2$; $j=1,2,3$).%
\footnote{Here positively and negatively charged $X_{W}^{i}$ are counted as one instance; they are taken as mass-degenerate and have the same decay branching ratios, but their production cross-sections are free parameters.} 
A priori they may be same-spin or opposite-spin partners to the SM particles.

\section{Simplified-model LHC results}

In order to confront individual \protomodels with a large number of LHC results, we make use of  \smo. 
Not employing any MC event simulation, this is computationally cheap, making it feasible to test hundreds of thousands of \protomodels within reasonable time. 
The experimental results stored in the \smo database fall into two main categories:
\begin{itemize}
	\item {\bf Upper Limit (UL)} results: these are the 95\% confidence level limits on the production cross sections (times BR) for simplified model topologies as a function of the BSM masses obtained by the experimental collaborations. 
	\item {\bf Efficiency Map (EM)} results: these correspond to signal efficiencies (more precisely, acceptance $\times$ efficiency, ${\cal{A}}\times \epsilon$, values) for simplified topologies as a function of the BSM masses for the  signal regions considered by the corresponding experimental analysis. 
\end{itemize}
The \smo database v1.2.4 used here includes results from 40 ATLAS and 46 CMS experimental searches at $\sqrt{s}=8$ and 13~TeV, corresponding to about 250 upper limit maps and 1,700 individual efficiency maps; see~\cite{Waltenberger:2020ygp} for details.

\section{Construction of a global likelihood}

\begin{figure}[h!]\centering
   \hspace*{-2mm}\includegraphics[width=0.43\textwidth]{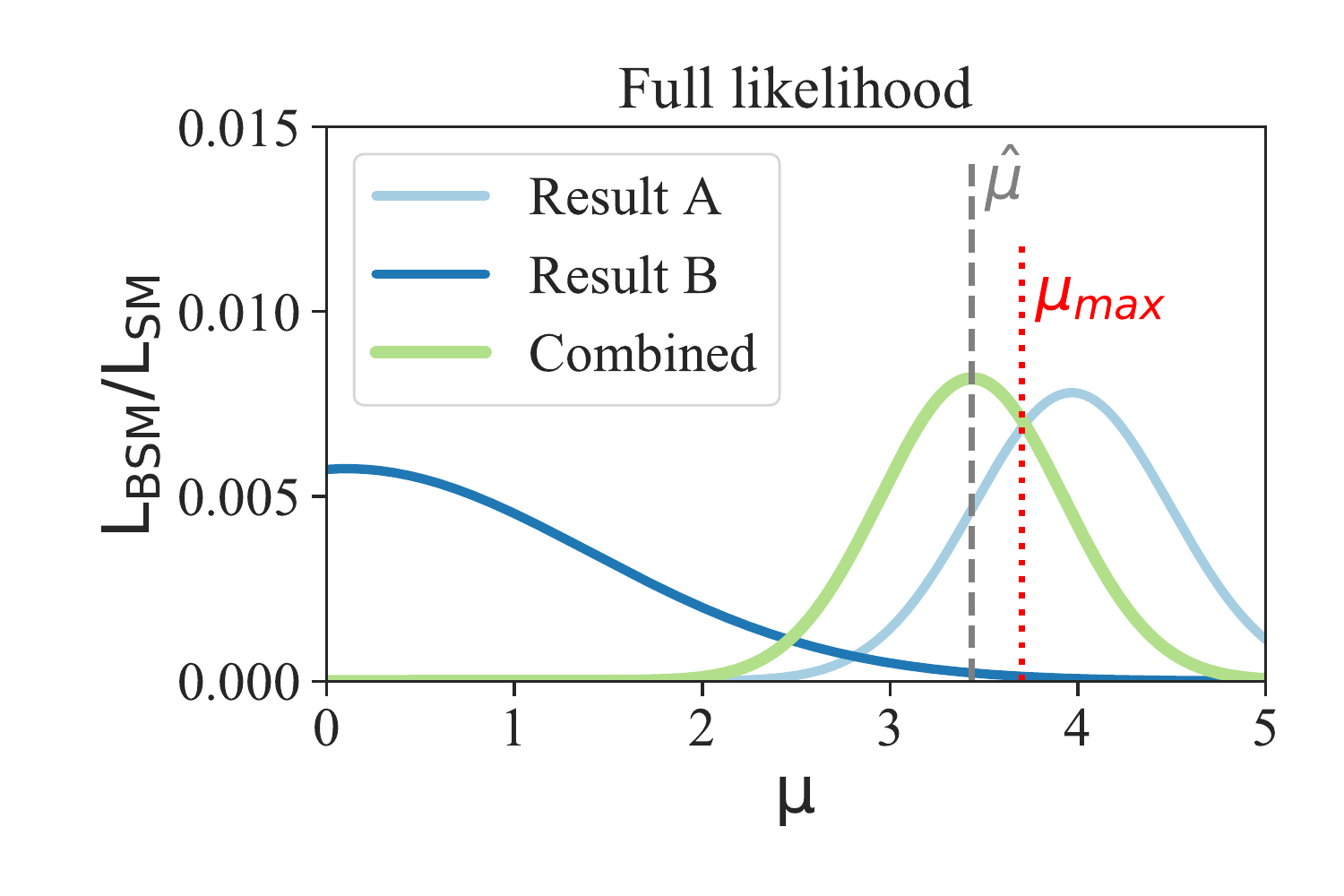}%
   \includegraphics[width=0.43\textwidth]{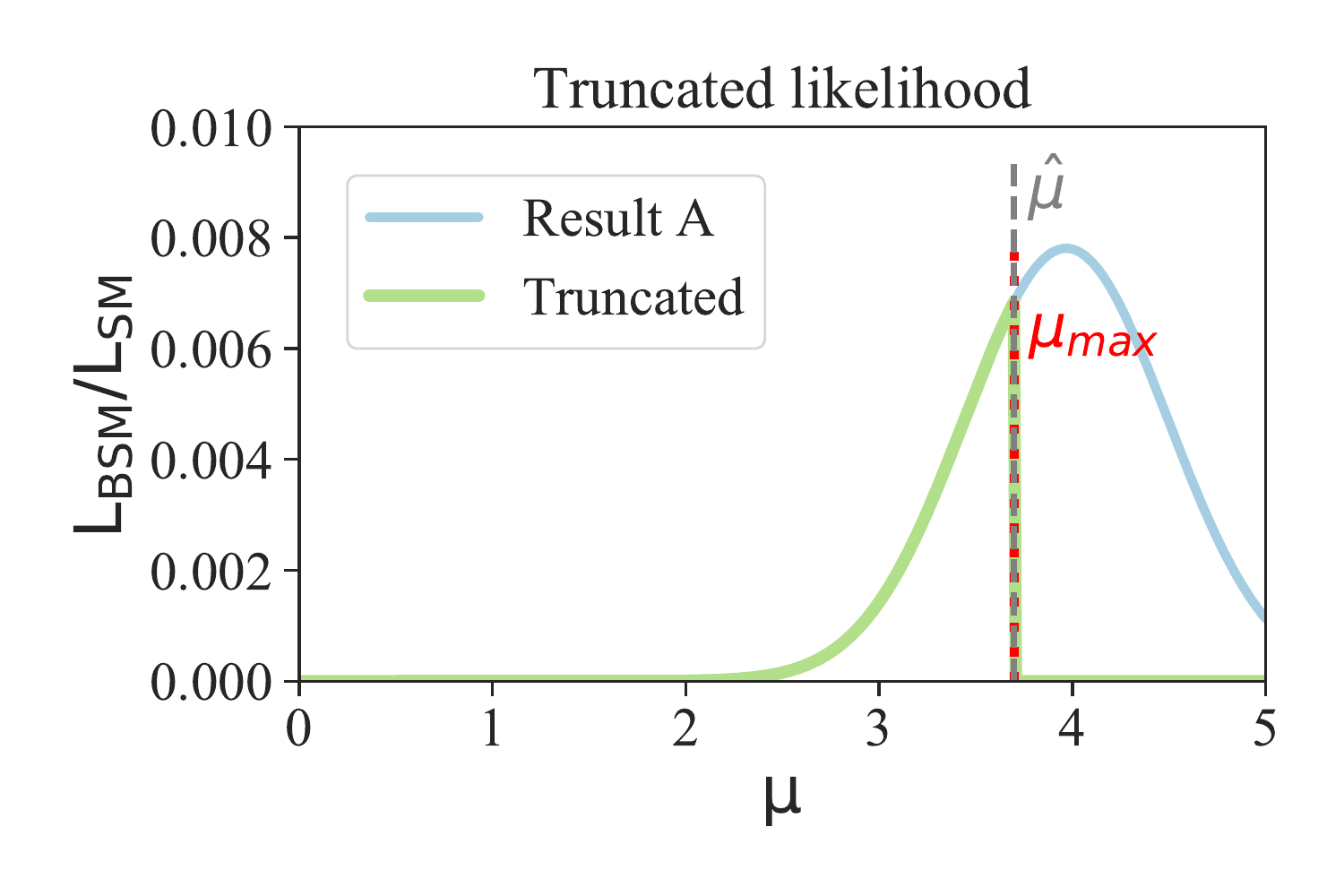}
   \vspace*{-4mm}
\caption{{\it Left}: an illustration of the signal likelihood as a function of the signal strength ($\mu$) for two results (A and B) and the corresponding combined likelihood. The value of $\hat{\mu}$ given by the maximum of the combined likelihood (green curve) is also shown as well as the upper limit on $\mu$ ($\mu_{max}$) given by Result B alone. {\it Right}: a similar example, but where it is not possible to combine both likelihoods, or the likelihood for Result B is not available. In this case the likelihood for Result A is truncated at $\mu_{max}$ (obtained from Result B). All the curves are normalized to the corresponding SM likelihood value ($\mu = 0$).}\label{fig:llhd}
\end{figure}

Key to our procedure is the construction of an approximate {\em global} likelihood for the signal, 
\begin{equation}
    \mathrm{L}_{\BSM}(\mu|D) =  P\left(D|\mu s + b + \theta \right) p(\theta) \,, \label{eq:globalL}
\end{equation}
which describes the plausibility of the signal strength $\mu$, given the data $D$. 
Here, $\theta$ denotes the nuisance parameters describing systematic
uncertainties in the signal ($s$) and background ($b$), while $p(\theta)$
corresponds to their probability distribution function.\footnote{Note that $s$, $b$ and $\theta$ in Eq.~\eqref{eq:globalL} are multi-dimensional quantities.}
The principle which we follow for combining individual likelihoods to a global one is illustrated in Figure~\ref{fig:llhd}. 

\subsection{Likelihoods for individual analyses}

The extent to which we can compute likelihoods for individual analyses crucially depends on the information available from the experimental collaborations. Simplified-model EMs allow us to determine the expected number of events for the hypothesised \protomodel. Together with the number of observed events, the expected backgrounds, and the uncertainties thereon, we can then construct a simplified likelihood by assuming a Gaussian distribution for the uncertainties and a Poissonian for the data; the nuisances are profiled over.   
This is a priori done per signal region. For some CMS analyses, a covariance matrix is available, 
which allows one to combine different signal regions, still in a simplified likelihood approach~\cite{simplifiedlikelihoods}. 
ATLAS has recently started to provide the full statistical models ~\cite{ATL-PHYS-PUB-2019-029} for some analyses, with which one can compute the likelihood at nearly~\footnote{Essentially up to small differences resulting from uncertainties in the EMs.} the same fidelity as in the experiment. 

If only ULs are available for an analysis, things look less good. 
If the expected ULs are available in addition to the observed ones, we can describe the likelihood as a truncated Gaussian, although this is often a crude approximation.\cite{Waltenberger:2020ygp}  
However, if only observed ULs are available, it is simply not possible to construct a reasonable likelihood function; such results can only be used to determine $\mu_{\rm max}$ for the critic.

\subsection{Combining analyses}\label{sec:combinations}

\begin{figure}[t!]\centering
   \includegraphics[width=0.42\textwidth]{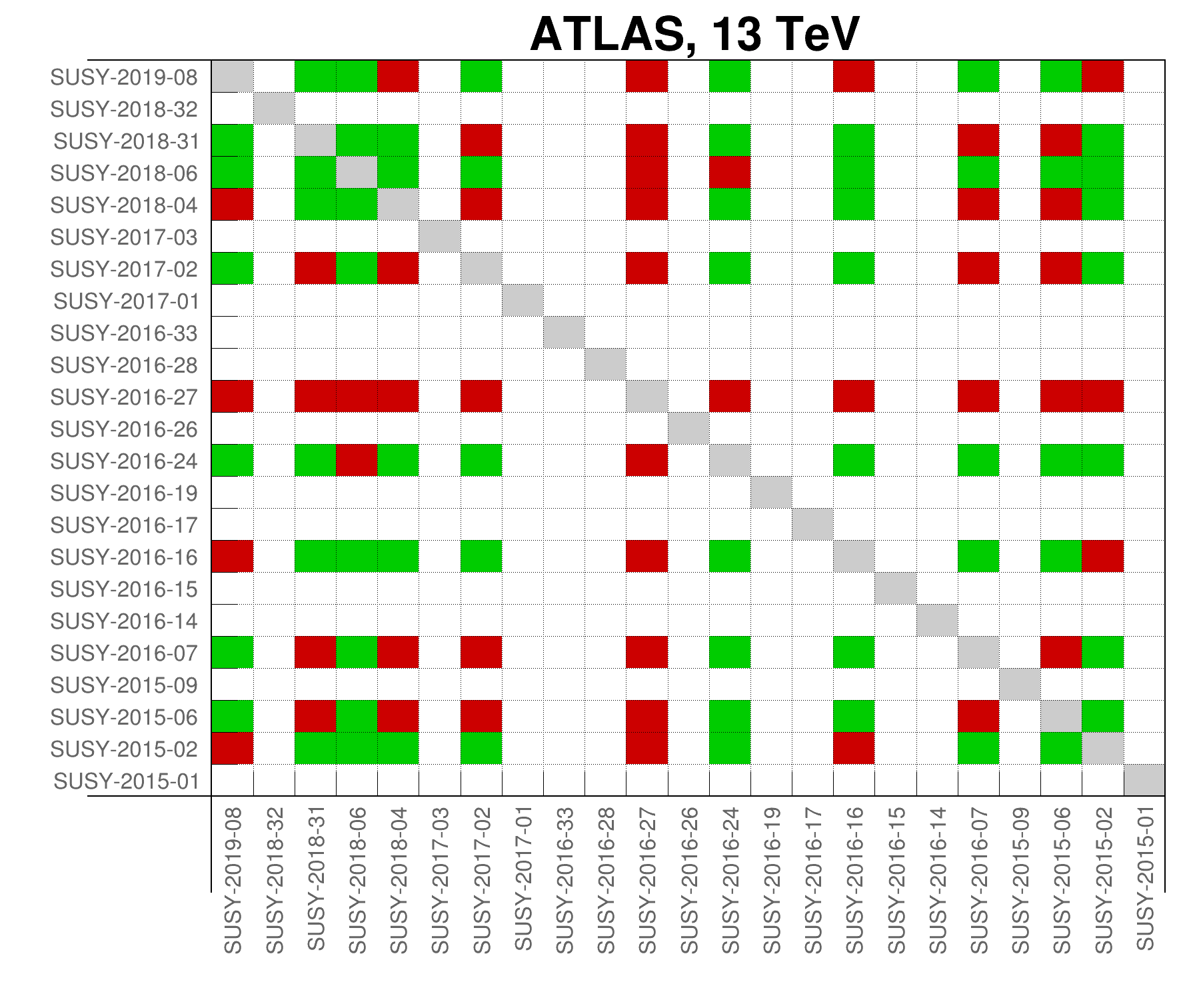}\quad%
   \includegraphics[width=0.42\textwidth]{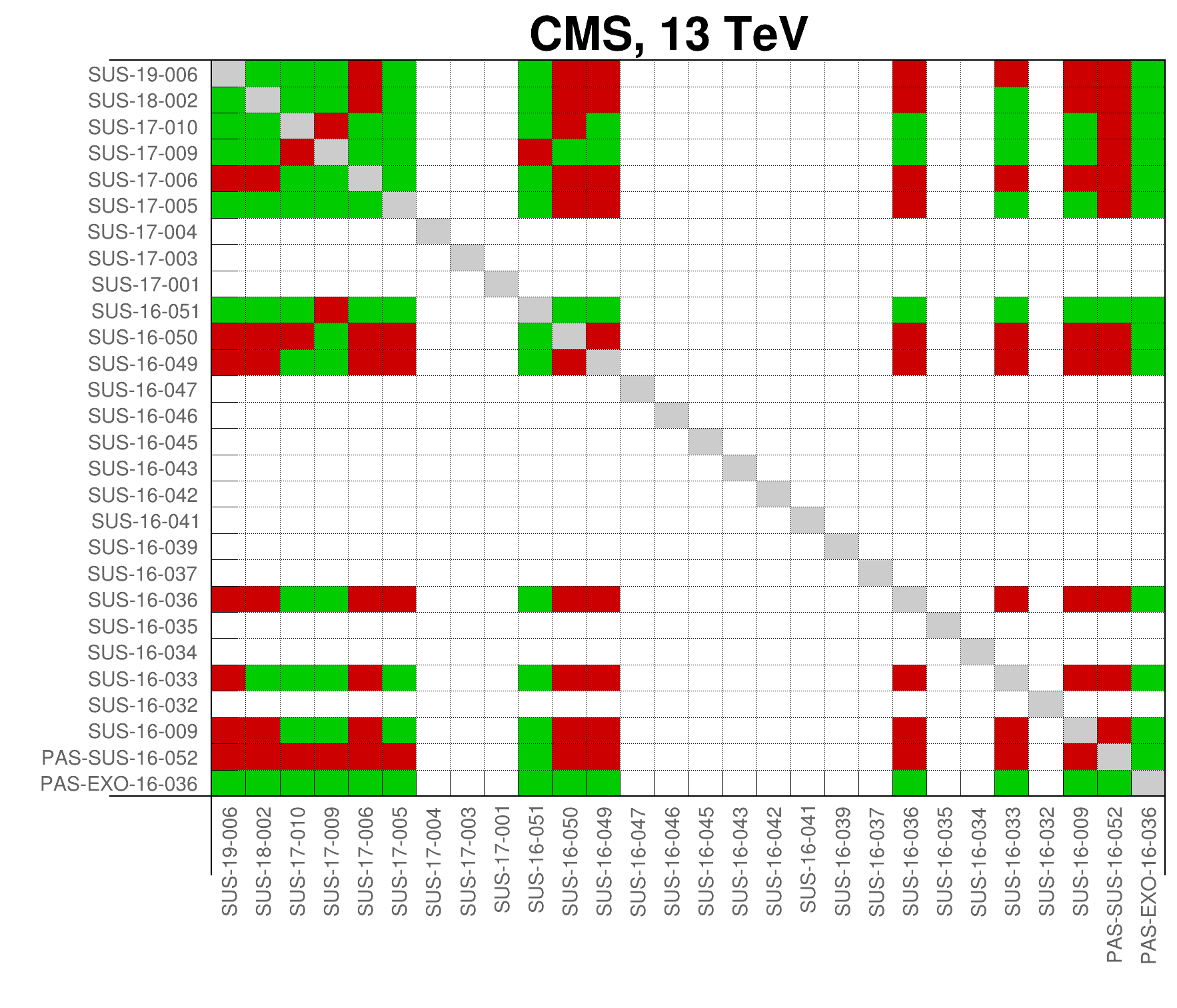}\vspace*{-2mm}
\caption{Binary correlation matrices for the 23 ATLAS (left) and 28 CMS (right) Run~2 analyses considered in this work, showing which ones are taken as approximately uncorrelated (green bins) and which are not (red bins). Analyses from different LHC runs or different experiments are always taken as uncorrelated. White bins denote analyses, for which no likelihood can be computed because only observed ULs are available.}\label{fig:corrmatrix}
\end{figure}

In principle, to combine the likelihoods from individual analyses, we would need to know their correlations. 
Lacking this information, we treat cross-analyses correlations in an approximate, binary way---a given pair of analyses is either considered to be approximately uncorrelated, in which case it may be combined (by multiplying the respective likelihoods), or it is not at all considered for combination.
We always assume results from different LHC runs and/or from different experiments to be approximately uncorrelated. Moreover, we also treat results with clearly different final states  
(e.g., fully hadronic final states vs.\ final states with leptons) to be uncorrelated. 
Figure~\ref{fig:corrmatrix} illustrates this for the Run~2 analyses considered in this work.

With these assumptions, we 
construct an approximate combined likelihood for subsets of LHC results, $\mathrm{L}_{\BSM}(\mu) = \prod_{i=1}^{n} \mathrm{L}_i(\mu)$, where the product is over all $n$ uncorrelated analyses and $\mu$ is the global signal strength. 
We refer to such subsets as {\it combinations} of results. Two important rules apply: 1.~any pair of results in the subset must be considered as uncorrelated; and 2.~any result which is allowed to be added to the combination, must be added. 
Information from all the other analyses, which are not included in the combination, is
accounted for as a constraint on the global signal strength $\mu$, i.e.\ $\mu_{\rm max}$.

A comment is in order at this point. Indeed, while a good number of analyses can be combined in the way described above, we also see from the white bins in Fig.~\ref{fig:corrmatrix}, that about half of the Run~2 analyses in the \smo v1.2.4  database consist of observed ULs only, and thus no proper likelihood can be computed for them. For Run~1, this concerns about one third of all analyses. Clearly, more EM-type results would be welcome to improve this situation.

\section{Test statistics}

Last but not least, in order to guide the MCMC-type walk and identify the \protomodels that best fit the data, we need a test statistic \K. While \K should increase for models 
which better satisfy all the constraints (which includes better fitting potential dispersed signals), it is also desirable to enforce the law of parsimony by reducing the test statistic of models with too many degrees of freedom. 
We define the test statistic as 
\begin{equation}
  \K := \max\limits_{\forall c \in C} \K^c \qquad{\rm with}\qquad 
  \K^{c} = 2 \ln \frac{
	\mathrm{L}^{c}_{\BSM}(\hat{\mu}) \cdot \pi(\BSM)}{\mathrm{L}^{c}_{\SM} \cdot \pi(\SM)} \,.
\label{eq:teststatistic}
\end{equation}

\noindent 
Here $\mathrm{L}^{c}_\BSM$ is the likelihood
for a combination $c$  of experimental results {\it given} the \protomodel, evaluated at the signal strength value $\hat{\mu}$, which maximizes the likelihood and satisfies $0 \leq \hat{\mu} < \mu_{max}$.  $\mathrm{L}^{c}_{\SM}$ is the corresponding SM likelihood, given by  $\mathrm{L}^{c}_\BSM (\mu=0)$. 
Finally, $\pi(\SM)$ and $\pi(\BSM)$ denote respectively the priors for the SM and the \protomodel. The total set of 
combinations of results, $C$, is determined as explained in Section~\ref{sec:combinations}. 

The \protomodel prior $\pi(\BSM)$ should penalize the test statistic for newly introduced particles, branching ratios, or signal strength multipliers, while $\pi(\SM) \equiv 1$. 
Here, we choose
\begin{equation}  
  \pi(\M) = \exp\left[- \left(\frac{n_\mathrm{particles}}{2} + \frac{n_\mathrm{BRs}}{4} + \frac{n_\mathrm{SSMs}}{8}\right) \right] \,,
\label{eq:prior}
\end{equation}
where $n_\mathrm{particles}$ is the number of new particles present in the \protomodel,  
$n_\mathrm{BRs}$ is the number of non-trivial branching ratios,\footnote{This means, for only one decay mode with 100\% BR,  $n_\mathrm{BRs}=0$.} and $n_\mathrm{SSMs}$ the number of signal strength multipliers.
This way, one particle with one non-trivial decay and two production modes is equivalent to one free parameter in the Akaike Information Criterion. For the SM, $n_\mathrm{particles}=n_\mathrm{BRs}=n_\mathrm{ssm}=0$, and $\pi(\SM) \equiv 1$.
The test statistic thus roughly corresponds to a $\Delta \chi^2$ of the \protomodel with respect to the SM, with a penalty for the new degrees of freedom. 

\section{Results}

\begin{figure}[t!]\centering
    \includegraphics[width=0.6\textwidth]{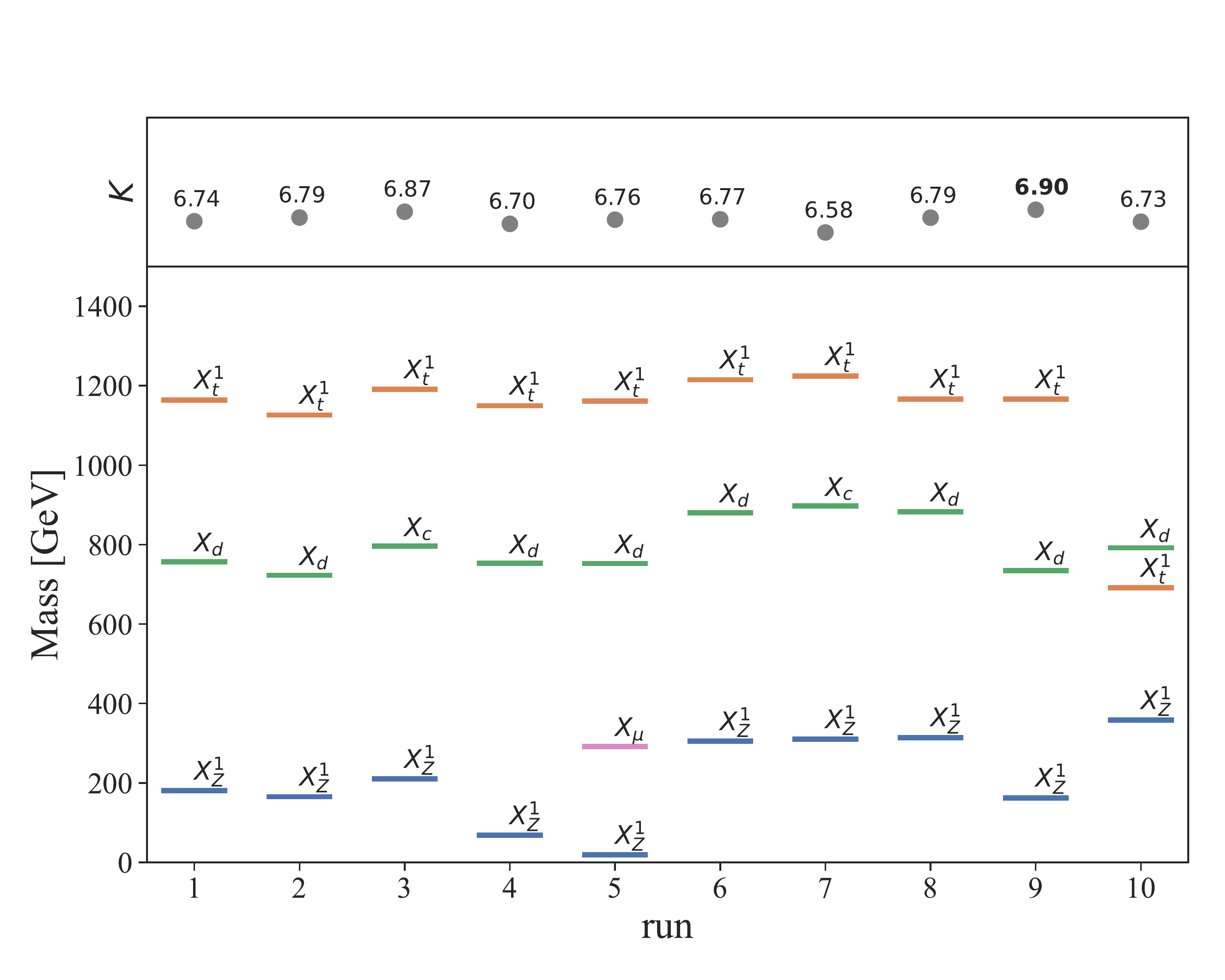}\vspace*{-2mm}		
		\caption{Particle content and masses for the \protomodels with highest test-statistic \K obtained in each of the 10 runs performed over the real database. The corresponding \K values are shown at the top. Note that for all practical purposes $X_d$ and $X_c$ are indistinguishable.}
		\label{fig:winners}
\end{figure}

For first physics results, we performed 10 runs of the walker algorithm, each employing $50$ walkers and $1,000$ steps/walker. Figure~\ref{fig:winners} displays the mass spectra of the \protomodels with the highest \K value from each run. 
Besides the $X_Z^1$ as the LBP, 
all models include one top-partner, $X_t^1$, and one light-flavor quark partner, $X_{d,c}$, 
and their test statistics are at $\K = 6.76 \pm 0.08$ 
thus showing the stability of the algorithm. 

For concreteness, 
let us take a closer look at the \protomodel with $\K = 6.9$ from run~9.\footnote{Note that the scenarios in Fig.~\ref{fig:winners} are driven by the same data and are statistically all equally plausible.} 
The $X_{d}$ with mass $\approx 700$~GeV is introduced in order to fit small excesses in the multijet+\met\xspace analyses from ATLAS at $\sqrt{s}= 8$ and  13~TeV~\cite{Aad:2014wea,Aaboud:2017vwy} and from CMS at $\sqrt{s}= 8$~TeV~\cite{Chatrchyan:2014lfa}. 
The $X_t^1$ with a mass of about $1.2$~TeV is introduced to fit  the $\approx 1.5\,\sigma$ and $2\,\sigma$ excesses observed in the 13 TeV CMS and ATLAS stop searches~\cite{Sirunyan:2017pjw,Aaboud:2017aeu}; 
indeed this is what mostly drives the \K value. 
Despite corresponding to small excesses, identifying the presence of such potential dispersed signals is one of the main goals of the algorithm presented here. The fact that these excesses appear in distinct
ATLAS and CMS analyses and 
can be explained by the introduction of a single top partner (actually with SUSY-like cross sections)
is another interesting outcome. 

Finally, we can also compute a global $p$-value for the SM  hypothesis by running over synthetic ``background-only'' data.  In the present setup, we find that $p({\rm SM})\approx 0.19$; see~\cite{Waltenberger:2020ygp} for details.

\section{Conclusions}

In view of the null results (so far) in the numerous  channel-by-channel searches for new particles, it becomes increasingly relevant to change perspective and attempt a more global approach to find out where BSM physics may hide. To this end, we presented a novel statistical learning algorithm that is capable of identifying potential dispersed signals in the slew of published LHC analyses. The task of the algorithm is to build candidate \protomodels from small excesses in the data, while at the same time remaining consistent with all other constraints. 
At present, this is based on the concept of simplified models, exploiting the \smo software framework and its large database of simplified-model results from ATLAS and CMS searches for new physics. \\

\noindent 
{\bf Acknowledgments:} S.K.\ was supported in part by the IN2P3 project ``Th\'eorie -- BSMGA''.

\section*{References}
\providecommand{\href}[2]{#2}\begingroup\raggedright\endgroup

\end{document}